\documentclass[12pt]{article}
\usepackage{amsmath}
\usepackage{graphicx}

\usepackage{amsfonts,amssymb,amsmath,amsthm}
\usepackage{hyperref}
\usepackage{graphicx}

\newcommand{\be}{\begin{equation}}
\newcommand{\ee}{\end{equation}}
\newcommand{\ba}{\begin{array}}
\newcommand{\ea}{\end{array}}
\newcommand{\bea}{\begin{eqnarray}}
\newcommand{\eea}{\end{eqnarray}}

\newcommand{\eps}{\epsilon}
\newcommand{\calH}{{\cal H }}
\newcommand{\calL}{{\cal L }}
\newcommand{\calM}{{\cal M }}
\newcommand{\calG}{{\cal G }}

\newcommand{\calE}{{\cal E }}

\newcommand{\calT}{{\cal T }}

\newcommand{\la}{\langle}
\newcommand{\ra}{\rangle}

\newcommand{\qqed}{\begin{flushright}$\qed$\end{flushright}}

\newtheorem{theorem}{Theorem}
\newtheorem{dfn}{Definition}
\newtheorem{lemma}{Lemma}
\newtheorem{corollary}{Corollary}
\newtheorem{claim}{Claim}

\title{Polynomial-time algorithm for simulation of weakly interacting quantum spin systems}
\author{Sergey Bravyi{\small${}^{(1)}$}\footnote{e-mail: sbravyi@us.ibm.com},\ \
David DiVincenzo{\small${}^{(1)}$}, and
Daniel Loss{\small${}^{(2)}$}\\ \\
{\it \small ${}^{(1)}$IBM T.J. Watson Research Center,
 Yorktown Heights NY 10598, USA.}\\
{\it \small ${}^{(2)}$Department of Physics, Klingelbergstrasse 82, University of Basel, 4056 Basel, Switzerland}
}

\begin{document}

\maketitle

\begin{abstract}
We describe an algorithm that computes the ground state energy and correlation functions
for $2$-local Hamiltonians in which interactions between qubits are  weak compared to
single-qubit terms.
The running time of the algorithm is polynomial in $n$ and $\delta^{-1}$, where
$n$ is the number of qubits, and $\delta$ is the required precision. Specifically, we consider Hamiltonians of the form
$H=H_0+\eps\, V$, where $H_0$ describes non-interacting qubits,
$V$ is a perturbation that involves arbitrary two-qubit interactions on a graph of
bounded degree, and $\eps$ is a small parameter. The algorithm works if $|\eps|$ is
below a certain threshold value $\eps_0$ that
depends only upon the spectral gap of $H_0$, the maximal degree of the
graph, and the maximal norm of the two-qubit interactions.
The main technical ingredient of the algorithm is  a generalized Kirkwood-Thomas
ansatz for the ground state. The parameters of the ansatz are computed using
perturbative expansions in powers of $\eps$. Our algorithm is closely related to
the coupled cluster method used in  quantum chemistry.
\end{abstract}

\newpage

\tableofcontents

\newpage

\section{Introduction and summary of results}
Perturbation theory provides a systematic way of getting approximations to
eigenvalues and eigenvectors for a variety of quantum spin models.
Arguably, a significant part of analytical and numerical results of  condensed matter physics
has been obtained using perturbative expansions in some small parameter.
Quite recently the methods of the perturbation theory have been successfully employed in
quantum complexity theory. In Ref.~\cite{KKR} Kempe, Kitaev, and Regev used
perturbative reductions to
show that  the problem of computing the ground state energy of a Hamiltonian with two-qubit
interactions is QMA-complete.  After that Terhal and Oliveira~\cite{OT} generalized this result
to local Hamiltonians on a 2D lattice.

Our main goal  is to examine whether the methods of the perturbation theory
provide an efficient computational algorithm for the simulation of quantum spin systems.
In this paper we focus on the simulation of low-temperature properties, namely
computing the ground state energy and spin-spin correlation functions for the ground state.
An efficient algorithm must have a running time $T=O(n^\alpha \, \delta^{-\beta})$, where
$n$ is the number of spins, $\delta$ is a precision up to which
we need to compute the ground state energy or a correlation function, and
$\alpha,\beta>0$ are some constants.

Before stating the results, let us describe the spin models that we shall consider.
Let $\calG=(\calL,\calE)$ be a graph with  a set of vertices $\calL$, $|\calL|=n$, and set of edges $\calE$.
Suppose $n$ spins-$1/2$ (qubits) are located at vertices $u\in \calL$ and
spin-spin interactions are located on edges $(u,v)\in \calE$.
The Hamiltonian is
\be
\label{Hmain}
H(\eps)=H_0+\eps\, V, \quad \quad  H_0=\sum_{u\in \calL} \Delta_u\, |1\ra\la1|_u, \quad \quad
V=\sum_{(u,v)\in \calE} V_{u,v}.
\ee
Here  $V_{u,v}$ is an arbitrary
operator acting on  a pair of qubits $u,v$, and $\eps$ is a real number.
The operators $H_0$ and $V$ are called the {\it unperturbed Hamiltonian} and the {\it perturbation}.
We shall always assume that $\Delta_u>0$ for all $u\in \calL$.
Accordingly, the unperturbed Hamiltonian  $H_0$ has a non-degenerate ground state
\[
|\Omega\ra=|0,0,\ldots,0\ra, \quad H_0\, |\Omega\ra=0.
\]
Most of the time, all we will need to know about $H_0$ and $V$ are the following parameters
\be
\Delta=\min_{u\in \calL} \Delta_u, \quad J=\max_{(u,v)\in \calE} \| V_{u,v}\|.
\ee
The parameter $\Delta$ is the gap between the smallest and the second smallest eigenvalue of $H_0$,
while the parameter $J$ characterizes a strength of the perturbation $V$.
Let $d$ be the maximum vertex degree of the graph $\calG$,
\be
d=\max_{u\in \calL} \left|\{ v\, : \, (u,v)\in \calE\}\right|.
\ee
The quantity we are interested in is the smallest eigenvalue of $H(\eps)$, which we
shall denote by $E(\eps)$. Clearly, $E(\eps)$ is a continuous concave function of $\eps$
and $E(0)=0$. Besides, since we assume that $\Delta>0$, the standard perturbation theory
arguments~\cite{Kato} show that $E(\eps)$ is analytic at $\eps=0$ and the Taylor series
\be
\label{Eseries}
E(\eps)=\sum_{p=1}^\infty E_p\, \eps^p
\ee
converges absolutely for $\|\eps V\| <\Delta/2$.
The following theorem proved by Yarotsky~\cite{Yarotsky} asserts that $E(\eps)$ is
non-degenerate for sufficiently small $\eps$ and sets a lower bound on the spectral gap.
\begin{theorem}
\label{thm:gap}
Suppose $|\eps|\le 2\eps_0$, where
\be
\eps_0=\frac{2^{-18}\,\Delta}{dJ}.
\ee
Then the smallest eigenvalue $E(\eps)$ has multiplicity $1$
and the gap between $E(\eps)$ and the second smallest eigenvalue of $H(\eps)$ is at least $\Delta/2$.
\end{theorem}
\noindent
(The explicit value of $\eps_0$ has not been stated in Ref.~\cite{Yarotsky}.)
We shall provide an alternative proof of Theorem~\ref{thm:gap} in Sections~\ref{sec:KT},\ref{sec:solution}.

As was shown by Osborne in Ref.~\cite{Osborne},
Theorem~\ref{thm:gap}  implies that expectation values of local observables  on the ground state
of $H(\eps)$  can be efficiently computed within any constant precision $\delta$
by simulating quantum adiabatic evolution along the path
connecting $H(0)$ and $H(\eps)$.  However,
the running time of such simulation scales
exponentially as a function of  $\delta^{-1}$.
As was noted in Ref.~\cite{Osborne}, it means that
simulation of  the adiabatic evolution
does not yield a polynomial-time algorithm  for computing the ground state energy.

The perturbation theory provides an approximation to the ground state energy by
truncating the series Eq.~(\ref{Eseries}) at sufficiently high order $p$.
In order to understand whether this approach can be used to construct an
efficient computational algorithm, two separate issues have to be
addressed:
\begin{enumerate}
\item[Q1:]  What is the convergence radius of the perturbative series?
\item[Q2:] What is the computational cost of finding the coefficients in the perturbative series?
\end{enumerate}
Note that the radius of convergence of the series Eq.~(\ref{Eseries}) is a property of the Hamiltonian
$H(\eps)$ only. It does not depend upon what particular perturbative expansion has been used to
find the coefficients $E_p$.  The following theorem allows one to answer the first question.
\begin{theorem}
\label{thm:conv}
The Taylor series $E(\eps)=\sum_{p=1}^\infty E_p\, \eps^p$ converges absolutely for
$|\eps|\le 2\eps_0$. Furthermore,
\be
\left|
E(\eps)-\sum_{q=1}^p E_q\,\eps^q\right| \le n\Delta 2^{-16-p} \quad \mbox{if} \quad |\eps|\le \eps_0.
\ee
\end{theorem}
Thus if one needs to compute $E(\eps)$ with a specified precision $\delta$, it suffices to
compute the coefficients $E_1,\ldots,E_p$, where $p=\log_2{(n\delta^{-1})}+O(1)$
(assuming that $\Delta$ is a constant that does not depend on $n$).

Answering the second question has nothing to do with the convergence radius of the
series Eq.~(\ref{Eseries}) (as long as it is non-zero). One can compute the coefficients $E_p$
by choosing $\eps$ so small that $\|\eps \, V\|\ll \Delta$. In this regime the standard perturbation
theory is applicable, for example, the self-energy operator formalism, see Refs.~\cite{KKR,AGD},
or the Rayleigh-Schr\"odinger expansion, see Ref.~\cite{Lindgren}.
Clearly, the computational cost of finding  the coefficients $E_p$ varies for different methods.

In the present paper we compute the coefficients $E_p$ using the Kirkwood-Thomas
ansatz for the ground state. It was  originally proposed in Ref.~\cite{KT83} for translation-invariant
Ising-like Hamiltonians with a transverse magnetic field. The translation-invariance
constraint has been removed in the later work by Datta and Kennedy~\cite{Kennedy}.
We use  the generalized Kirkwood-Thomas ansatz proposed by Yarotsky~\cite{Yarotsky}
which is applicable to any spin Hamiltonian with sufficiently weak interactions.
It allows us to prove the following.
\begin{theorem}
\label{thm:algorithm}
Suppose $d$ is a fixed constant independent of $n$. Then
there exists an algorithm  with a running time
$n\exp{(O(p))}$  that takes as input a triple
$(H_0,V,p)$ and outputs $E_1,\ldots,E_p$.
\end{theorem}
An immediate consequence of Theorems~\ref{thm:conv},\ref{thm:algorithm} is
\begin{corollary}
Suppose $\Delta$, $J$, $d$ are fixed constants independent of $n$ and
$|\eps|\le \eps_0$. Then
there exists an  algorithm with a running time $poly(n,\delta)$
that computes $E(\eps)$ with an absolute error at most $\delta$.
\end{corollary}
Besides, it follows from Theorems~\ref{thm:conv},\ref{thm:algorithm}
that the energy {\it density} $E(\eps)/n$ can be computed with  a precision
$\delta$ in a time $n \cdot poly(\delta^{-1})$.

Note that while computing the coefficients $E_1,\ldots,E_p$ we cannot afford the running time
to grow faster than $\exp{(O(p))}$ (for fixed $n$) since we need $p\sim \log{(n\delta^{-1})}$
to achieve the desired accuracy. The perturbative expansion based on the Kirkwood-Thomas ansatz
has two special features that make the scaling $\exp{(O(p))}$ possible:
(i) The parameters of the ansatz are complex amplitudes $C(M)$ assigned to subsets
of vertices $M\subseteq \calL$.
The recursive equations specifying the amplitudes  $C(M)$ are described by
a polynomial of a {\it constant degree}, see Section~\ref{subs:fpowers};
(ii) The perturbative expansion $C(M)=\sum_{p=1}^\infty C_p(M)\,\eps^p$
has a property known as the linked cluster theorem, namely, $C_p(M)=0$ unless
$M$ can be covered by a connected subgraph of size $O(p)$, see Section~\ref{subs:lct}.
The number of such subgraphs grows only exponentially with $p$, see Section~\ref{subs:upper}.
It implies that the number of non-zero coefficients $C_p(M)$ grows as $n\exp{(O(p))}$,
see Section~\ref{subs:C_p(M)}.

Naturally, one could run the algorithm from Theorem~\ref{thm:algorithm}
to compute the truncated series for $E(\eps)$
even if $|\eps|>\eps_0$.
The running time will be polynomial in $n$ and $\delta^{-1}$ as long as $|\eps|$ is smaller
than the convergence radius $R$ of the series Eq.~(\ref{Eseries}). Although we believe that $R$
 must be close to $\Delta/(dJ)$ (see a discussion at Section~\ref{sec:discussion}),
its exact value cannot be easily found.
In practical simulations, one could evaluate $R$ by computing
sufficiently many coefficients $E_p$ and using the fact that
$R^{-1}$ is the largest accumulation point of a sequence $|E_p|^{1/p}$, $p=1,\ldots,\infty$,
see Ref.~\cite{Lang}. Note that in general the singular point (points) of $E(\eps)$ with $|\eps|=R$
does not lie on the real axis and thus cannot be identified with a quantum phase transition point
of $H(\eps)$ (since we consider finite systems, the latter is not even well defined).

Obviously, efficient computation of $E(\eps)$ is possible due to the presence of a small
parameter $\eps$ in the problem. However it should be emphasized that the condition $|\eps|\le \eps_0$
does not imply that the ground state $|\psi\ra$ of $H(\eps)$ is close to the ground state $|\Omega\ra$ of the unperturbed
Hamiltonian $H_0$. In fact, one should expect that $|\psi\ra$ and $|\Omega\ra$ are almost orthogonal for
large $n$\footnote{This effect is analogous to the well-known ``orthogonality catastrophe'' observed
by Anderson in Ref.~\cite{OC} for non-interacting fermions in a presence of a scattering potential.}.
 To illustrate this statement, consider as an example the perturbation $V=-J \sum_{u\in \calL} X_u$,
where $X$ is the Pauli $\sigma^x$ operator, and the unperturbed Hamiltonian $H_0=\Delta \sum_{u\in \calL}
|1\ra\la 1|_u$.  Clearly, the ground state of  $H(\eps)$ is a product of one-qubit
states, $|\psi\ra=\bigotimes_{u\in \calL} |\psi_u\ra$. A simple calculation shows that
$\la 0|\psi_u\ra =\cos{(\theta/2)}$, where $\cos{(\theta)}=(1+4\eps^2 J^2/\Delta^2)^{-1/2}$.
Thus for any fixed $\eps$ the overlap $\la \Omega|\psi\ra=(\cos{(\theta/2)})^n$ gets exponentially
small as $n$ increases. However the reduced density matrices of the ground states
$|\psi\ra$ and $|\Omega\ra$ for any subset of qubits of constant size are indeed close to each other for small $\eps$.
In other words, for small $\eps$ the state $|\psi\ra$ describes
small density quantum fluctuations of the background state $|\Omega\ra$.
One could speculate that this statement remains true for arbitrary weak perturbations as well.
The Kirkwood-Thomas ansatz for the ground state  of $H(\eps)$ used in the present paper provides a convenient way
to quantify the ``density of quantum fluctuations" and prove that it is indeed small
for $|\eps|\le \eps_0$. Unfortunately, our approach does not allow us  to make any statements about the validity of the
area law or to decide whether the ground state can be well approximated using the PEPS ansatz,
see Ref.~\cite{PEPS}.

The Kirkwood-Thomas ansatz is not well suited for computing spin-spin correlation functions because it
provides an unnormalized ground state. We avoid this problem using the standard relation between
the correlation functions and the linear response of the ground state energy to a small perturbation.
It allows us to prove
\begin{theorem}
\label{thm:corr}
Let $O_{u,v}$ be a Hermitian operator acting non-trivially only on qubits $u,v\in \calL$.
Suppose $\|O_{u,v}\|\le 1$. The expectation value of $O_{u,v}$ on the
ground state of $H(\eps)$ 
can be computed with a precision $\delta$ in a time $T=poly(\delta^{-1})$ as long as
$|\eps|\le \eps_0/2(d+1)$.
\end{theorem}
\noindent
{\it Remark:} After the present work has been completed, it was communicated to us by
F.~Verstraete~\cite{Verstraete:pc} that the simulation algorithm based on the Kirkwood-Thomas ansatz
is closely related to the coupled cluster method
originally introduced by Coester~\cite{Coester58}. The coupled cluster method is extensively used
for numerical simulations in quantum chemistry, see a review~\cite{Crawford99},
as well as in condensed matter physics, see a review~\cite{Bishop06} and the references therein.
Accordingly, from the perspective of practical simulations, the algorithm described in the
present paper is certainly not a new one.
However, we believe that our results provide the first rigorous proof that the coupled cluster method
yields a polynomial-time simulation algorithm for spin Hamiltonians with weak interactions.

The rest of the paper is organized as follows.
Section~\ref{sec:KT} provides the necessary background on the generalized Kirkwood-Thomas
ansatz. It mostly follows Ref.~\cite{Yarotsky}, although some of our proofs
are technically different (in particular, Lemma~\ref{lemma:gap}).
Section~\ref{sec:solution} shows how to solve the
Kirkwood-Thomas equations using a power series and proves Theorem~\ref{thm:conv}.
In Section~\ref{sec:LCT} we prove  that our perturbative expansion obeys the well-known
linked cluster theorem and  establish an upper bound on the
number of linked clusters on a graph. The algorithms for computing the ground state
energy and spin-spin correlation functions are explicitly described in Section~\ref{sec:algorithms}
which provides a proof of Theorems~\ref{thm:algorithm},\ref{thm:corr}.
Some open problems are discussed in Section~\ref{sec:discussion}.
Appendix~A contains a technical lemma proving submultiplicativity
of the norm of creation operators.

\section{Kirkwood-Thomas ansatz for the ground state}
\label{sec:KT}

\subsection{Creation operators}
Define one-qubit  operator $a^\dag=|1\ra\la 0|$. Let $a_u^\dag$ be the operator $a^\dag$ on qubit $u$ tensored with the
identity on all other qubits. For any non-empty subset of vertices $M\subseteq \calL$ denote
$a_M^\dag=\prod_{u\in M} a_u^\dag$. Note that the operators
$a_M^\dag$ are nilpotent, $(a_M^\dag)^2=0$, and that they pairwise commute: $a_M^\dag a_K^\dag=
a_K^\dag a_M^\dag$.
Also, one can easily check that the  operators $\{a_M^\dag\}$, $\emptyset\ne M\subseteq \calL$
are linearly independent.
(All these definitions and properties apply to $a$ and $a_M$ operators as well).

\begin{dfn}
A creation operator is an operator that can be written as
\[
C=\sum_{\emptyset\ne M\subseteq \calL} C(M)\, a_M^\dag
\]
for some complex numbers $C(M)$.
\end{dfn}
For any given creation operator $C$ the coefficients $C(M)$ are uniquely defined by $C(M)=\la \Omega|a_M\, C|\Omega\ra$.
\begin{claim}
\label{claim:1}
Any state $|\psi\ra$ satisfying $\la \Omega|\psi\ra=1$ can be uniquely written as
$|\psi\ra=exp{(-C)} \, |\Omega\ra$ for some creation operator $C$.
\end{claim}
Remark: the exponent above is defined by its Taylor series. The nilpotence of operators $a_M^\dag$
implies that $C^k=0$ for any $k$ greater than the number of qubits $n=|\calL|$,
 so the Taylor series can be truncated at $k=n$.\\
{\bf Proof:}
Clearly, the states $\{ a_M^\dag\, |\Omega\ra\}$, $M\subseteq \calL$, constitute the orthonormal
basis of the $n$-qubit Hilbert space.
Let $|\psi\ra=\sum_{M\subseteq \calL} \psi(M) a_M^\dag \, |\Omega\ra$. Equation $|\psi\ra=\exp{(-C)}\, |\Omega\ra$
is equivalent to a system of equations
\be
\label{psi(C)}
\psi(\emptyset)=1, \quad C(M)=-\psi(M) + \sum_{k=2}^{|M|}  \frac{(-1)^k}{k!} \sum_{M=M_1\cup \ldots \cup M_k}
C(M_1)\cdots C(M_k), \quad M\subseteq \calL, \quad M\ne \emptyset.
\ee
Here the second summation is over all partitions of $M$ into $k$ disjoint non-empty sets $M_1,\ldots,M_k$.
Suppose we have already found all coefficients $C(M)$ with $|M|\le p$. Then Eq.~(\ref{psi(C)})
assigns a unique value to all coefficients $C(M)$ with $|M|=p+1$.
Thus the system Eq.~(\ref{psi(C)}) has a unique solution.
\qed

\subsection{Ansatz for the ground state}
Our goal is to find an eigenvector $|\psi\ra$
satisfying $H(\eps)\, |\psi\ra= E(\eps) \, |\psi\ra$, where $E(\eps)$ is the smallest eigenvalue of $H(\eps)$.
We shall use the following ansatz for $|\psi\ra$ (we don't care about the normalization):
\be
\label{gs}
|\psi\ra = \exp{(-C)}\, |\Omega\ra,  \quad  C=\sum_{\emptyset \ne M\subseteq \calL} C(M)\, a_M^\dag.
\ee
Claim~\ref{claim:1} asserts  that the ground state can be represented in this form unless it is
orthogonal to $|\Omega\ra$. Since we don't require $|\psi\ra$ to be a normalized state, the ansatz Eq.~(\ref{gs})
is meaningful only if $C$ is a bounded operator.
We shall define a norm of a creation operator as
\be
\label{norm}
\|C\|_1 = \max_{u\in \calL} \sum_{M\ni u} |C(M)|.
\ee
Thus the ansatz  Eq.~(\ref{gs}) must be supplemented by a requirement that  $C$ is a creation operator with
a finite norm $\|C\|_1$.
Of course, it may happen that $H$ has several eigenvectors of the form Eq.~(\ref{gs}).
One has to  invoke some extra arguments to select an eigenvector corresponding to the smallest eigenvalue, see
 subsection~\ref{subs:select}.

The ``physical meaning" of the norm $\|C\|_1$ can be illustrated by considering a product state:
$|\psi\ra=\bigotimes_{u\in \calL} |\psi_u\ra$, where $|\psi_u\ra=|0\ra + \alpha_u\, |1\ra$.
Obviously, $|\psi\ra=\exp{(-C)}\, |\Omega\ra$ with $C=-\sum_{u\in \calL} \alpha_u\, a_u^\dag$.
Accordingly, $\|C\|_1=\max_{u\in \calL} |\alpha_u|$.  Thus one can think about  $\|C\|_1$ as
a density of quantum fluctuations.

Using the identity $\exp{(C)} \exp{(-C)}=I$ valid for arbitrary operator $C$, see~\cite{Bhatia}, one can rewrite
the Schr\"odinger equation $H(\eps)\, |\psi\ra = E(\eps)\, |\psi\ra$ as
\be
\label{SE}
\exp{(\hat{C})} (H_0)|\Omega\ra + \eps\, \exp{(\hat{C})}(V) |\Omega\ra = E(\eps)\, |\Omega\ra.
\ee
Here we introduced a superoperator\footnote{In our context a superoperator is a linear operator acting
on the space of linear operators on $\calH$.
Throughout the paper we shall use a notation $\hat{A}$ for a superoperator
$\hat{A}(X)=AX-XA$ associated with a linear operator $A$.} $\hat{C}$ such that
\[
\hat{C}(X)=CX-XC.
\]
The exponent $\exp{(\hat{C})}$ is defined by the Taylor series.
The advantage of the ansatz Eq.~(\ref{gs}) is that we can truncate expansion of the exponent
$\exp{(\hat{C})}$ after a few lowest orders since all higher order terms turn out to be identically zero.
It follows from the two lemmas stated below.
\begin{lemma}
\label{lemma:H_0}
Let $C_1$, $C_2$ be creation operators. Then
\be
\hat{C}_1 \hat{C}_2 (H_0)=0.
\ee
\end{lemma}
\noindent
{\bf Proof:} To simplify notations we shall consider operators $a$ instead of $a^\dag$.
Let $u,M_1,M_2\subseteq \calL$ and  $X=\hat{a}_{M_1} \hat{a}_{M_2} (|1\ra\la 1|_u)$. By linearity, it is enough to
prove that $X=0$. Since the operators $a_{M_1}$ and $a_{M_2}$ commute, $X=0$ unless
$u\in M_1\cap M_2$. Then $[a_{M_2},|1\ra\la 1|_u]=a_{M_2}$ and $X=[a_{M_1},a_{M_2}]=0$.
\qqed

\begin{lemma}
\label{lemma:V}
Let $C_1,C_2,\ldots,C_5$ be creation operators. Then
\be
\hat{C}_1 \hat{C}_2 \cdots \hat{C}_5 (V)=0.
\ee
\end{lemma}
{\bf Proof:}
 To simplify notations we shall consider operators $a$ instead of $a^\dag$.
Let $M_1,M_2,\ldots,M_5\subseteq \calL$, $(u,v)\in \calE$ and
$X=\hat{a}_{M_1}\,\hat{a}_{M_2}\cdots \hat{a}_{M_5}(V_{u,v})$.
By linearity it is enough to prove that $X=0$.
Since the operators $a_{M_1},\ldots,a_{M_5}$ commute with each other,
 $X=0$ unless each of the subsets $M_j$ contains at least one of the vertices $u,v$.
Therefore, expanding the commutators one can represent $X$ as a linear combination of $2^5$ terms,
where each term contains at least five operators $a_u$, $a_v$ on the pair of qubits $u,v$.
Some of these operators $a$ are on the right of $V_{u,v}$ and some of them are on the left.
Thus at least three operators $a$  are on the same side of $V_{u,v}$.
Then at least two operators $a$ act on the same side of $V_{u,v}$ and on the same qubit. Thus
each of the $2^5$ terms in $X$ contains either $a_u^2$ or $a_v^2$. Thus
 $X=0$.
\qqed

Combining Lemmas~\ref{lemma:H_0},\ref{lemma:V} we get the following truncations:
\be
\label{te1}
\exp{(\hat{C})} (H_0)= H_0 + \hat{C}(H_0),
\ee
\be
\label{te2}
\exp{(\hat{C})} (V)=\sum_{k=0}^4 \frac1{k!} \hat{C}^k (V).
\ee
Here a convention $\hat{C}^0(V)=V$ is adopted.

Let us point out an analogy between the truncation effect observed above and the Lieb-Robinson
bound~\cite{LRbound1,LRbound2}. The latter asserts that for any local observable $O_u$ acting only on a qubit $u$
and for any Hamiltonian $H$ with short-range interactions of bounded norm the time evolved observable
$O_u(t)=\exp{(i\hat{H}t)}(O_u)$ can be approximated very well by an operator acting only on
spins within distance $v|t|$ from $u$, where $v$ is a group velocity. If one takes a creation operator $C$
for which the coefficients $C(M)$ are non-zero only for subsets $M$ of size $O(1)$ (an analogue of short-range
interactions), then the "time-evolved" observable $\exp{(\hat{C})}(O_u)$ acts only on the spins
within distance $O(1)$ from $u$ (apply the same arguments as in the proof of Lemma~\ref{lemma:V}).
As opposed to the Lieb-Robinson bound scenario, the size of a region acted on by the evolved operator
does not depend on the norm of $C$ (which is analogous to the evolution time )
and no approximations are involved.

\subsection{Kirkwood-Thomas equations}
Substituting Eqs.~(\ref{te1}) into the  Schr\"odinger equation Eq.~(\ref{SE}) and taking into
account that $H_0\, |\Omega\ra=0$ one gets
\be
\label{SE1'}
-\sum_{\emptyset\ne M\subseteq \calL} C(M) H_0\,  a_M^\dag \, |\Omega\ra + \eps  \exp{(\hat{C})}(V) \, |\Omega\ra = E(\eps) \, |\Omega\ra.
\ee
Let us introduce eigenvalues of the unpertubed Hamiltonian $E_0(M)$ such that
\be
\label{E_0}
H_0 \, a_M^\dag \, |\Omega\ra=
E_0(M) \, a_M^\dag \, |\Omega\ra, \quad
E_0(M)=\sum_{u\in M} \Delta_u.
\ee
Multiplying Eq.~(\ref{SE1'})  on the left by $\la \Omega|a_M$, $M\ne \emptyset$, and employing Eq.~(\ref{te2}) one arrives at
\be
\label{SE2}
C(M)=\frac{\eps}{E_0(M)}  \sum_{k=0}^4 \frac1{k!} \la \Omega| a_M \hat{C}^k (V)|\Omega\ra, \quad \emptyset \ne M\subseteq \calL.
\ee
Following~\cite{Yarotsky}, we shall refer to Eq.~(\ref{SE2}) as {\it  Kirkwood-Thomas equations}.
Similarly, multiplying Eq.~(\ref{SE1'}) by $\la \Omega|$ on the left one gets
\be
\label{SE3}
E(\eps)=\eps\sum_{k=0}^4 \frac1{k!} \la \Omega| \hat{C}^k (V)|\Omega\ra.
\ee
It is clear that the Kirkwood-Thomas equations Eq.~(\ref{SE2}) may have several
solutions $C$ since the equations do not explicitly include the eigenvalue $E(\eps)$. In the worst
case when neither eigenvector of $H(\eps)$  is orthogonal to $|\Omega\ra$
the Kirkwood-Thomas equations would have $2^n$ solutions since any eigenvector could be
represented in the form Eq.~(\ref{gs}).
We shall explain how to select the solution corresponding to the smallest eigenvalue in the next
subsection.

The following lemma asserts that the norm $\|\cdot\|_1$ has a property analogous to
submultiplicativity.  It is the main technical tool that allows one to manipulate easily with equations like Eq.~(\ref{SE2}).
\begin{lemma}
\label{lemma:norms}
Let $k$ be any integer and $C_1,\ldots,C_k$ be creation operators. Define a creation operator $C$ such that
\[
C=\sum_{\emptyset\ne M\subseteq \calL} C(M) a_M^\dag
\quad \mbox{where}    \quad
C(M)=\frac1{E_0(M)} \la \Omega|a_M \hat{C}_1\cdots \hat{C}_k(V)|\Omega\ra.
\]
Then
\be
\label{main1}
\|C\|_1\le \frac{2^{13} dJ}{\Delta} \prod_{j=1}^k \|C_j\|_1.
\ee
Besides,
\be
\label{main2}
|\la \Omega|\hat{C}_1\cdots \hat{C}_k(V_{u,v})|\Omega\ra | \le 2^4 J \prod_{j=1}^k \|C_j\|_1
\quad \mbox{for any} \quad (u,v)\in \calE.
\ee
\end{lemma}
The proof of the lemma is presented in Appendix~A.

\subsection{A lower bound on the spectral gap}
\label{subs:select}
Suppose we can find {\it some} eigenvalue $E'(\eps)$ of the Hamiltonian $H(\eps)$ such that $E'(0)=0$,
$E'(\eps)$ is a continuous function of $\eps$,
and $E'(\eps)$ has multiplicity $1$ for $|\eps|\le \eps_c$. Then it follows immediately that
$E'(\eps)$ is the {\it smallest} eigenvalue of $H(\eps)$ for all $|\eps|\le \eps_c$.
Of course, the main difficulty in using this argument is proving non-degeneracy of an eigenvalue.
The following lemma asserts that a solution of the Kirkwood-Thomas equations
Eq.~(\ref{SE2}) with a sufficiently small norm $\|C\|_1$ corresponds to a non-degenerate eigenvalue
separated from the rest of the spectrum by a constant gap.

\begin{lemma}
\label{lemma:gap}
Suppose $H(\eps)\,|\psi\ra =E(\eps)\, |\psi\ra$, where
 $|\psi\ra =\exp{(-C)}\, |\Omega\ra$
and $C$ is a creation operator with a finite norm $\|C\|_1$ satisfying the inequality
\be
\label{gap_bound}
1> \frac{2^{14} dJ |\eps|}{\Delta} \sum_{k=0}^3 \frac{(\|C\|_1)^k}{k!}.
\ee
Then  $E(\eps)$ has multiplicity $1$ and any other eigenvalue of $H(\eps)$
is separated from $E(\eps)$ by a gap at least $\Delta/2$.
\end{lemma}

{\bf Proof:}
Let us abbreviate $H\equiv H(\eps)$.
Assume that
$H\, |\phi\ra= (E(\eps)+\delta)\, |\phi\ra$ where $|\delta|<\Delta/2$ and the states
$|\psi\ra$, $|\phi\ra$ are linearly independent (the latter condition is fulfilled automatically if $\delta\ne 0$).
We can always write  $|\phi\ra$ as
\be
\label{phiaux}
\exp{(C)}\, |\phi\ra=\sum_{M\subseteq \calL} B(M)\, a_M^\dag \, |\Omega\ra
\ee
for some complex numbers $B(M)$.  Note that  $B(M)\ne 0$ for some non-empty set $M$
since otherwise $|\phi\ra$ is proportional to $|\psi\ra$.
Thus we can define a creation operator $B=\sum_{\emptyset \ne M\subseteq \calL} B(M)\, a_M^\dag$
with a non-zero norm $\|B\|_1>0$.
Using commutativity $[C,B]=0$ we can represent $|\phi\ra$ as
\[
|\phi\ra=B\, |\psi\ra + B(\emptyset) \, |\psi\ra.
\]
Then the eigenvalue equations $H\, |\phi\ra = (E(\eps)+\delta)\, |\phi\ra$ and
$H\, |\psi\ra = E(\eps)\, |\psi\ra$ imply
\be
\label{long}
[B,H]\, |\psi\ra = [B,H-E(\eps) I]\, |\psi\ra = B (H-E(\eps) I )\, |\psi\ra - (H-E(\eps)I )\, |\phi\ra =
-\delta\, |\phi\ra = -\delta B\, |\psi\ra - \delta B(\emptyset) \, |\psi\ra.
\ee
Commutativity $[C,B]=0$  yields $\exp{(\hat{C})}(B)=B$. Hence, multiplying Eq.~(\ref{long})
by $\exp{(C)}$ on the left one arrives at
\be
\label{Baux}
[B, \exp{(\hat{C})}(H)]\, |\Omega\ra + \delta B\, |\Omega\ra + \delta B(\emptyset) \, |\Omega\ra =0.
\ee
From Lemma~\ref{lemma:H_0} we know  that $[B,\exp{(\hat{C})}(H_0) ]=[B,H_0]$.
Choosing any $M\ne \emptyset$ and multiplying Eq.~(\ref{Baux}) by $\la \Omega|a_M$ on the left one gets
\[
B(M) = \frac{\eps}{E_0(M)-\delta} \la \Omega |a_M \hat{B} \exp{(\hat{C})}(V) |\Omega\ra=\frac{\eps E_0(M)}{E_0(M)-\delta}
\sum_{k=0}^3
\frac1{k!}\frac1{E_0(M)}
\la \Omega|a_M \hat{B}\hat{C}^k(V)|\Omega\ra.
\]
Here we have taken into account that $\hat{B}\hat{C}^k(V)=0$ for $k\ge 4$, see Lemma~\ref{lemma:V}.
Note that condition $|\delta|<\Delta/2$ implies a bound $|E_0(M)/(E_0(M)-\delta)|\le 2$.
Applying  Lemma~\ref{lemma:norms} to the operator $B$ and using the triangle inequality for the norm one gets
\[
\|B\|_1 \le \frac{2^{14} d J |\eps|}{\Delta} \|B\|_1 \sum_{k=0}^3 \frac1{k!} (\|C\|_1)^k.
\]
Since $\|B\|_1>0$ we can divide both sides by $\|B\|_1$ getting an inequality
opposite to the one stated in the lemma.  Thus the assumption from which we started the proof
leads to a contradiction.
\qqed

{\it Remark: } Note that at $\eps=0$ the Hamiltonian $H(\eps)=H_0$ has many degenerate eigenvalues,
so one can certainly find two eigenvalues with separation $|\delta|<\Delta/2$. It might seem to be in
contradiction with the lemma above. However at $\eps=0$ the condition that $\|C\|_1$ is finite can not
be fulfilled for degenerate eigenvalues, since the corresponding eigenvectors are orthogonal to $|\Omega\ra$.

\begin{corollary}
\label{corollary:gap}
Suppose $H(\eps)$
has an eigenvector $|\psi\ra=\exp{(-C)}\, |\Omega\ra$ with an
eigenvalue $E'(\eps)$ such that $E'(\eps)$ is a continuous function of $\eps$,
$E'(0)=0$, and $\|C\|_1\le c_{max}$ for all $|\eps|\le \eps_{c}'$.  Define $\eps_{c}''$ such that
\[
1 = \frac{2^{14} d J \eps_{c}''}{\Delta} \sum_{k=0}^3 \frac1{k!} (c_{max})^k.
\]
Let $\eps_c=\min{(\eps_{c}',\eps_{c}'')}$.
Then for all $|\eps|\le \eps_c$\\
(1) $E'(\eps)$ is the smallest eigenvalue of $H(\eps)$\\
(2) $E'(\eps)$ has multiplicity $1$\\
(3) $E'(\eps)$ is separated from the rest of the spectrum by a gap at least $\Delta/2$
\end{corollary}
{\bf Proof:} (1) Indeed, Lemma~\ref{lemma:gap} implies that no level crossings involving the eigenvector $|\psi\ra$
can occur for $|\eps|\le \eps_c$. Since $|\psi\ra$ is the ground state for $\eps=0$, it is the ground state
for all $|\eps|\le \eps_c$. (2) and (3) follow immediately from Lemma~\ref{lemma:gap}.
\qqed

\section{Solution of the Kirkwood-Thomas equations}
\label{sec:solution}

In their original paper~\cite{KT83} Kirkwood and Thomas employed the expansion in powers of $\eps$
in order to find the ground state.
Alternative approach proposed by Datta and Kennedy in Ref.~\cite{Kennedy} and generalized by
Yarotsky in Ref.~\cite{Yarotsky}
is to regard equation Eq.~(\ref{SE2}) as a fixed point equation for a non-linear map on the space
of creation operators. One can prove that this map is a contraction in the unit ball  (for a properly defined metric)
if $\eps$ is below certain threshold value. Then one can invoke Brouwer fixed point theorem
to argue that the unit ball contains a unique fixed point. Although the latter method is more elegant, we
adopt the original Kirkwood-Thomas approach based on power series, because it naturally lends itself
for getting approximation to the ground state energy with a controllable error.

\subsection{Solution by formal power series}
\label{subs:fpowers}
Let us first solve Kirkwood-Thomas equation Eq.~(\ref{SE2}) in terms of formal power series ignoring the convergence issue.
Recall that $C=\sum_{M\subseteq \calL} C(M) \, a_M^\dag$,
where the sum is over all non-empty sets.
Define a series
\be
\label{Cseries}
C(M)=\sum_{p=1}^\infty C_p(M)\, \eps^p, \quad \emptyset \ne M \subseteq \calL.
\ee
Let us agree that $C_0(M)=0$ for any $M$.
Define also
\be
\label{Cseries1}
C_p=\sum_{\emptyset \ne M\subseteq \calL} C_p(M)\, a_M^\dag, \quad
\hat{C}_p= \sum_{\emptyset \ne M\subseteq \calL} C_p(M)\, \hat{a}_M^\dag,
\ee
so that $C=\sum_{p=1}^\infty C_p\, \eps^p$ and $\hat{C}=\sum_{p=1}^\infty \hat{C}_p \, \eps^p$.
Substituting the series Eqs.~(\ref{Cseries},\ref{Cseries1}) into the Kirkwood-Thomas equation Eq.~(\ref{SE2}) and equating
the coefficients for each power of $\eps$ one gets
\be
\label{Cformal1}
C_1(M) = E_0(M)^{-1}\, \la \Omega| a_M\, V|\Omega\ra,
\ee
\be
\label{Cformal2}
C_p(M)= E_0(M)^{-1}\sum_{k=1}^4 \frac1{k!} \sum_{p_1+ \ldots + p_k=p-1}
\la \Omega| a_M \, \hat{C}_{p_1} \cdots \hat{C}_{p_k}(V)|\Omega\ra,
\quad p\ge 2.
\ee
Clearly, the equations above have a unique solution.
Substituting Eqs.~(\ref{Cseries},\ref{Cseries1}) into the formula for the ground state energy Eq.~(\ref{SE3}) one gets
\be
\label{Eformal}
E(\eps)=\sum_{p=1}^\infty E_p \, \eps^p,
\quad
E_1=\la \Omega|V|\Omega\ra, \quad
E_p =\sum_{k=1}^4 \frac1{k!} \sum_{p_1+\ldots +p_k =p-1}
\la \Omega|\hat{C}_{p_1} \cdots \hat{C}_{p_k}(V)|\Omega\ra, \quad p\ge 2.
\ee
Of course, formal power series do not represent an actual solution of the Kirkwood-Thomas equations
unless we prove their convergence.

\subsection{Convergence of $C$-series}
\label{subs:Cconv}
We would like to prove that the series $C=\sum_{p=1}^\infty C_p\, \eps^p$
are convergent
with respect to the norm Eq.~(\ref{norm}) with a non-zero convergence radius. We shall need to get
a lower bound on the convergence radius  in terms of $\Delta$, $d$, and $J$.
Clearly, it is enough to analyze convergence of  the series
\be
\label{chiseries}
\chi(\eps)=\sum_{p=1}^\infty \chi_p\, \eps^p, \quad \chi_p=\|C_p\|_1.
\ee
Note that $\|C\|_1\le \chi(|\eps|)$.
\begin{lemma}
\label{lemma:chi}
The series $\chi(\eps)=\sum_{p=1}^\infty \chi_p\, \eps^p$ converges absolutely for
\be
|\eps|\le 2\eps_0=\frac{2^{-17} \Delta}{dJ}.
\ee
Besides, for any $\eps$ as above one has the following bounds:
\be
\label{mupupper}
|\chi(\eps)|\le 2^{-15}, \quad
\chi_p\le \frac{2^{-15}}{(2\eps_0)^p} \quad \mbox{for} \quad p\ge 1.
\ee
\end{lemma}
{\bf Proof:}
Let us first get an upper bound on $\|C_1\|_1$.
From  Eq.~(\ref{Cformal1}) it clear that $C_1(M)=0$ unless $M\subseteq \{u,v\}$ for
some edge $(u,v)\in \calE$. Let $u\in \calL$ be the vertex that achieves the maximum in
$\|C_1\|_1=\max_{u\in \calL} \sum_{M\ni u} |C_1(M)|$.
Then the sum contains at most $d+1$ sets $M$, namely, $M=\{u\}$ and $M=\{u,v\}$ for
$(u,v)\in \calE$. Therefore $\|C_1\|_1\le (d+1)J/\Delta \le 2dJ/\Delta$, that is
\be
\label{chi1}
\chi_1\le \frac{2dJ}{\Delta}.
\ee
Define a polynomial function $F_p$ of real variables $x_1,\ldots,x_{p-1}$  according to
\be
\label{Fp}
F_p(x_1,\ldots,x_{p-1}) = x_{p-1} + \frac12
\sum_{p_1+p_2=p-1} x_{p_1} x_{p_2} +
\frac16 \sum_{p_1+p_2+p_3=p-1} x_{p_1}x_{p_2} x_{p_3}
+ \frac1{24} \sum_{p_1+\ldots + p_4=p-1} x_{p_1} x_{p_2} x_{p_3} x_{p_4}.
\ee
Applying Lemma~\ref{lemma:norms} and triangle inequality to Eq.~(\ref{Cformal2}) one gets
\be
\label{chip}
\chi_p\le \frac{2^{13} dJ}{\Delta} F_p(\chi_1,\ldots,\chi_{p-1}), \quad p\ge 2.
\ee
To simplify notations, define constants
\be
a=\frac{2dJ}{\Delta}, \quad b=\frac{2^{13} dJ}{\Delta},
\ee
so that $\chi_1\le a$ and $\chi_p\le bF_p(\chi_1,\ldots,\chi_{p-1})$ for $p\ge 2$.
Consider the formal power series
\be
\label{museries}
\mu(\eps)=\sum_{p=1}^\infty \mu_p \eps^p,
\quad
\mu_1=a,
\quad
\mu_p=b F_p(\mu_1,\ldots,\mu_{p-1}), \quad p\ge 2.
\ee
Since the polynomial $F_p$ has non-negative coefficients one can prove inductively  that
$\chi_p\le \mu_p$ for all $p\ge 1$. Hence it suffices to prove that the series Eq.~(\ref{museries})
converges absolutely for $|\eps|\le 2\eps_0$.

Our strategy will be to guess a  function $\mu(\eps)$ analytic for $|\eps|\le 2\eps_0$
whose Taylor series at $\eps=0$ coincides with the series Eq.~(\ref{museries}).
By inspecting the recursive relation Eq.~(\ref{museries}) one can easily
convince oneself that $\mu(\eps)$ has to obey the following equation
\be
\label{muan}
\mu(\eps) =a\, \eps +
b\, \eps \left(
\mu(\eps) + \frac12 \mu^2(\eps) + \frac16 \mu^3(\eps) + \frac1{24} \mu^4(\eps)\right).
\ee
We can use it to write down the inverse function
\be
\label{inverse}
\eps(\mu) = \frac{\mu}{Q(\mu)}, \quad Q(\mu)=a + b\left(\mu + \frac12 \mu^2 + \frac16 \mu^3
+ \frac1{24}\mu^4\right).
\ee
Simple algebra shows that
\be
\label{Q>}
|Q(\mu)|\ge \frac{a}2 \quad \mbox{if} \quad |\mu|\le \frac{a}{4b}=2^{-14}.
\ee
Thus $\eps(\mu)$ is analytic  for $|\mu|\le 2^{-14}$. Define a set $M=\{\mu\, : \, |\mu|\le 2^{-15}\}$.
\begin{claim}
\label{claim:bijection}
Let $\eps$ be a complex number such that $|\eps|\le 2\eps_0$. Then
equation $\eps(\mu)=\eps$ has a unique solution $\mu\in M$.
\end{claim}
{\bf Proof:}
One can easily show that for any $\mu_1,\mu_2\in M$
\be
\label{Lcondition}
|Q(\mu_1)-Q(\mu_2)|\le 2b |\mu_1-\mu_2|.
\ee
Assume $\eps(\mu_1)=\eps(\mu_2)=\eps$ for some $\mu_1,\mu_2\in M$. If $\eps=0$
then $\mu_1=\mu_2=0$. Assume $\eps\ne 0$. Then $\mu_1,\mu_2\ne 0$ and
\[
\mu_1-\mu_2=\frac{\mu_2 (Q(\mu_1)-Q(\mu_2))}{Q(\mu_2)}.
\]
Applying the lower bound Eq.~(\ref{Q>}) and the upper bound Eq.~(\ref{Lcondition}) we get
\[
|\mu_1-\mu_2|\le \frac{2^{-15} |Q(\mu_1)-Q(\mu_2)|}{a/2} \le \frac{2^{-13}\, b|\mu_1-\mu_2|}{a} \le \frac12 |\mu_1-\mu_2|.
\]
Thus $\mu_1=\mu_2$ and equation $\eps(\mu)=\eps$ has at most one solution $\mu\in M$.
Therefore $\eps\, : \, M \to \eps(M)$ is an injection. Let us prove that $\eps(M)$ contains a ball
of radius $2\eps_0$. Indeed, $\eps(M)$ is an open set and $0\in \eps(M)$. Let $\gamma$ be the boundary of $M$,
i.e., a circle of radius $2^{-15}$ centered at $0$. Then $\eps(\gamma)$ is the boundary of $\eps(M)$.
For any $\mu \in \gamma$ one has $|Q(\mu)|\le a+ 2b|\mu| =a+ 2^{-14}b \le 2a$.
Thus $\eps(M)$ contains a ball of radius
\[
R=\min_{\mu \in \gamma} |\eps(\mu)| \ge \frac{2^{-15}}{2a} =\frac{2^{-17} \Delta}{d J}=2\eps_0
\]
It completes the proof of the claim.

Let $K=\{\eps\, : \, |\eps|\le 2\eps_0\}$.
From Claim~\ref{claim:bijection} we infer that $\eps(\mu)$ is an analytic bijection from
the set $\eps^{-1}(K)\subseteq M$ to the set $K$. It follows from the inverse function
theorem for analytic functions, see~\cite{Lang}, that the inverse function $\mu(\eps)$
is analytic for $\eps\in K$.   Therefore the series Eq.~(\ref{museries}) converges absolutely for $|\eps|\le 2\eps_0$.

The upper bound on $\mu_p$ can be obtained by standard methods using Cauchy's formula:
\[
\mu_p=\frac1{2\pi i} \oint_{|\eps|=2\eps_0}  \frac{\mu(\eps) d\eps}{\eps^{p+1}}.
\]
Thus
\[
|\mu_p|\le \frac1{(2\eps_0)^p} \max_{\eps\, : \, |\eps|=2\eps_0} |\mu(\eps)|\le \frac{2^{-15}}{(2\eps_0)^p}.
\]
Recall that $\chi_p\le \mu_p$, so the lemma is proved.
\qqed

One can summarize the results of this subsection as follows.
\begin{corollary}
\label{corollary:Cnorm}
Suppose $|\eps|\le 2\eps_0$. Then
the Kirkwood-Thomas equations Eq.~(\ref{SE2}) have
a unique solution $C$ defined by the power series
Eq.~(\ref{Cseries}) with
$\|C\|_1\le 2^{-15}$.
\end{corollary}

\subsection{Convergence of $E$-series}
\label{subs:Econ}
In this subsection we analyze convergence of the series $E(\eps)=\sum_{p=1}^\infty E_p\, \eps^p$
for the eigenvalue obtained from the Kirkwood-Thomas equation, see Eq.~(\ref{Eseries}).
 \begin{lemma}
\label{lemma:chi1}
The series $E(\eps)=\sum_{p=1}^\infty E_p\, \eps^p$ converges absolutely for
\be
|\eps|\le 2\eps_0=\frac{2^{-17} \Delta}{dJ}.
\ee
Besides,
\be
\label{Eppupper}
|E_p|\le \frac{2^{-16}\, n\Delta}{(2\eps_0)^p} \quad \mbox{for} \quad p\ge 1.
\ee
\end{lemma}
{\bf Proof:}
Applying Lemma~\ref{lemma:norms} and triangle inequality to Eq.~(\ref{Eformal}) one gets
\be
|E_1|\le ndJ, \quad |E_p| \le 2^4 ndJ F_p(\chi_1,\ldots,\chi_{p-1}), \quad p\ge 2,
\ee
where $\chi_p=\|C_p\|_1$ and the polynomial $F_p$ is defined in Eq.~(\ref{Fp}).
Define a formal series
\be
e(\eps)=\sum_{p=1}^\infty e_p\, \eps^p,
\quad
e_1=ndJ, \quad e_p=2^4 ndJ F_p(\chi_1,\ldots,\chi_{p-1}), \quad p\ge 2.
\ee
By definition, $|E_p|\le e_p$ for all $p$.
Besides,  $e(\eps)$ can be expressed in terms of $\chi(\eps)=\sum_{p=1}^\infty \chi_p\, \eps^p$ as
\[
e(\eps) =2^4 ndJ \eps \left(
\chi(\eps) + \frac12 \chi^2(\eps) + \frac16 \chi^3(\eps) + \frac1{24} \chi^4(\eps)
\right) + ndJ\, \eps.
\]
This equality can be verified by equating coefficients for each power of $\eps$.
Lemma~\ref{lemma:chi} implies that $\chi(\eps)$ is analytic for $|\eps|\le 2\eps_0$.
Therefore $e(\eps)$ and $E(\eps)$ are analytic for $|\eps|\le 2\eps_0$ and the first statement of the lemma
is proved.
In order to get an upper bound on $e_p$ (and thus on $E_p$), use Cauchy's formula:
\[
e_p=\frac1{2\pi i}\oint_{|\eps|=2\eps_0} \frac{ e(\eps) d\eps }{\eps^{p+1}}.
\]
It follows from Lemma~\ref{lemma:chi} that $|\chi(\eps)|\le 2^{-15}$ for
$|\eps|\le 2\eps_0$.
Therefore
\[
|e_p|\le \frac1{(2\eps_0)^p} \max_{\eps\, : \, |\eps|=2\eps_0} |e(\eps)| \le \frac1{(2\eps_0)^p}
\left( 2^4 ndJ (2\eps_0) 2^{-14} + ndJ(2\eps_0) \right) \le \frac{2^{-16}n \Delta}{(2\eps_0)^p}.
\]
The lemma is proved.
\qqed
\begin{corollary}
\label{corollary:Delta/2}
Suppose $|\eps|\le 2\eps_0$. Then the series Eq.~(\ref{Eformal})
converges absolutely to the smallest eigenvalue of $H(\eps)$.
The smallest eigenvalue is non-degenerate and is separated from
the rest of the spectrum by a gap at least $\Delta/2$.
\end{corollary}
{\bf Proof:} It follows from Corollary~\ref{corollary:gap}. Indeed, we have already shown
that the conditions of Corollary~\ref{corollary:gap} are satisfied with $\eps_c'=2\eps_0$
and $c_{max}=2^{-15}$, see Corollary~\ref{corollary:Cnorm}.
It yields $\eps_c''\le 2^{-15} \Delta/(dJ)$. Thus $\eps_c=\min{(\eps_c',\eps_c'')}=2\eps_0$.
\qqed

Lemma~\ref{lemma:chi1}  allows one to estimate an error resulting from truncation of  the series for the ground state
energy at a finite order $p$.
\begin{corollary}
\label{corollary:error}
Suppose $|\eps|\le \eps_0$. Then
\be
\left|
E(\eps) - n\sum_{q=1}^p E_q\, \eps^q
\right|
\le n \Delta 2^{-16-p}.
\ee
\end{corollary}
{\bf Proof:} Use Eq.~(\ref{Eppupper}).
\qqed
Summarizing, we have proved Theorems~\ref{thm:gap},\ref{thm:conv}.

\section{Linked cluster theorems}
\label{sec:LCT}
Throughout this section we shall use a term {\it linked cluster} which refers to a subset
of vertices inducing a connected subgraph of $\calG$. More formally,
\begin{dfn}
A subset $M\subseteq \calL$ is called a linked cluster iff for any $u,v\in \calM$
there exists a sequence of vertices $u_0,u_1,\ldots,u_t\in M$ such that
$u_0=u$, $u_t=v$ and
$(u_j,u_{j+1})\in \calE$ for all $j=0,\ldots,t-1$.
\end{dfn}
\begin{dfn}
A connected size of a subset $M\subseteq \calL$ is the minimal size of a linked cluster
that contains all vertices of $M$. We shall denote a connected size of $M$ as $|M|_c$.
\end{dfn}

\subsection{Linked cluster expansion for the ground state}
\label{subs:lct}
\begin{lemma}
\label{lemma:lcprop}
Let $C(M)=\sum_{p=1}^\infty C_p(M) \eps^p$ be the solution of the Kirkwood-Thomas equations
obtained in Section~\ref{sec:solution}. Then
\be
\label{lcprop}
C_p(M)=0 \quad \mbox{unless} \quad |M|_c\le p+1.
\ee
\end{lemma}
{\bf Proof:}
We shall prove the lemma by induction in $p$.
From Eq.~(\ref{Cformal1}) one infers that $C_1(M)=0$ unless $M\subseteq \{u,v\}$ for some
edge $(u,v)\in \calE$. In particular, $C_1(M)=0$ unless $|M|_c\le 2$.
It proves the statement of the lemma for $p=1$.
Suppose  the statement is proved for $p=1,\ldots,q-1$.
From Eq.~(\ref{Cformal2}) one infers that $C_q(M)$ is a linear combination of terms like
\be
\label{matrixe}
x=C_{p_1}(M_1) \cdots C_{p_k}(M_k) \la \Omega|a_M \hat{a}^\dag_{M_1} \cdots \hat{a}^\dag_{M_k}(V_{u,v})|\Omega\ra,
\ee
where $p_1+\ldots+p_k=q-1$.
Let us figure out under what circumstances the matrix element in Eq.~(\ref{matrixe}) can be non-zero.
\begin{claim}
\label{claim:x}
Let $M,M_1,\ldots,M_k\subseteq \calL$ be non-empty sets,   $N=M_1\cup \ldots \cup M_k$,
$(u,v)\in \calE$. Denote
\[
y=\la \Omega| a_M \, \hat{a}^\dag_{M_1} \cdots \hat{a}^\dag_{M_k} (V_{u,v})|\Omega\ra.
\]
Then $y=0$ unless the following conditions are met:\\
(i) Each set $M_1,\ldots, M_k$ contains at least one of the vertices $u,v$.\\
(ii) $N\backslash \{u,v\} \subseteq M \subseteq N \cup \{u,v\}$.
\end{claim}
Remark: this claim is true even for $M=\emptyset$ if one adopts a convention $a_\emptyset=I$.

\noindent
{\bf Proof:} Suppose some set $M_j$ contains neither $u$ nor $v$.  Then $a_{M_j}^\dag$
commutes with $V_{u,v}$ as well as with all operators $a_{M_i}^\dag$ for $i\ne j$. Thus $y=0$.
Suppose condition $N\backslash \{u,v\} \subseteq M$ is violated.
Then there exists a set $M_j$ and a vertex $w\in M_j$ such that $w\ne u,v$ and $w\notin M$.
Thus $a_{M_j}^\dag$ contains a factor  $a_w^\dag$ which commutes with all other
operators involved in $y$. By moving $a_w^\dag$ leftwards one can show that
each of $2^k$ terms in $y$ starts from $\la \Omega|a_w^\dag$, that is, $y=0$.
Suppose condition $M \subseteq N \cup \{u,v\}$ is violated, that is,
there exists a vertex $w\in M$ such that $w\notin N$ and $w\ne u,v$.
Then the operator $a_M$ contains a factor $a_w$ which commutes with all other
operators involved in $y$. By moving $a_w$ rightwards one can show that
each of $2^k$ terms in $y$ tails with $a_w\, |\Omega\ra$, that is, $y=0$.
\qqed
\noindent
Returning to Eq.~(\ref{matrixe}) we conclude that
 $x=0$ unless each set $M_j$ contains either $u$ or/and $v$,
and $M\subseteq N \cup \{u,v\}$. Let $\tilde{M}_j$ be a linked cluster of minimal size
containing $M_j$, that is, $|M_j|_c=|\tilde{M}_j|$. Let $\tilde{N}= \tilde{M}_1\cup \ldots \cup \tilde{M}_k$
and $C=\tilde{N} \cup \{u\} \cup \{v\}$. Then $C$ is a linked cluster and $M\subseteq C$.
Note that
\[
|C|\le \sum_{j=1}^k |\tilde{M_j}| +2 -k=\sum_{j=1}^k |M_j|_c +2 -k
\]
where we have taken into account that each $\tilde{M}_j$ contains either $u$ or/and $v$.
By induction hypothesis we have $C_{p_j}(M_j)=0$ unless $|M_j|_c\le p_j+1$. Thus
for any non-zero term $x$ one has
\[
|C|\le \sum_{j=1}^k (p_j +1) +2 -k = q-1 + k + 2 - k=q+1.
\]
Thus $C_q(M)=0$ unless $|M|_c\le q+1$.
\qed

\subsection{Upper bound on the number of linked clusters}
\label{subs:upper}
The following lemma asserts that the number of linked clusters of size $p$ containing a given
vertex grows at most exponentially with $p$ (if the maximal degree of the graph $d$ is a constant).
To the best of our knowledge, this lemma has been originally proved in Ref.~\cite{AGP07} by
Aliferis, Gottesman, and Preskill in the context of fault-tolerant quantum computation\footnote{The authors
became aware of it after completion of the present work.}.
\begin{lemma}
\label{lemma:counting}
Let $N_p(u)$ be the number of linked clusters with $p$ vertices that contain a vertex $u$
and $N_p=\max_{u\in \calL} N_p(u)$. Then
\be
N_p\le (4d)^{p-1}.
\ee
\end{lemma}
\noindent
{\bf Proof:}
Let $\calT_p(u)$ be a set of trees with $p$ vertices that contain a vertex $u$
(naturally, we consider only those trees that are subgraphs of $\calG$).
Let $T_p(u)=|\calT_p(u)|$ be the number of such trees.
For any tree $T\in \calT_p(u)$, a set of vertices of $T$ is a linked cluster that contains $u$.
Conversely, if $M\ni u$ is a linked cluster, $|M|=p$, consider a subgraph $G_M$ induced by $M$.
Then any spanning tree of $G_M$ belongs to $\calT_p(u)$.
Thus  $N_p(u)\le T_p(u)$.

Denote $T_p=\max_{u\in \calL} T_p(u)$.
Obviously, $T_1=1$ and $T_2\le d$. Let us prove that
\be
\label{Tbound1}
T_p\le d \sum_{p_1+p_2=p} T_{p_1} T_{p_2},
\ee
where the convention $T_0=0$ is adopted.
Indeed, for any edge $e$ incident to a vertex $u$ define a set $\calT_p(u,e)$
that includes all trees $T\in \calT_p(u)$ that contain an edge $e$.
Let $T_p(u,e)=|\calT_p(u,e)|$.
 Clearly,
\be
\label{Tbound2}
\calT_p(u)=\cup_e \calT_p(u,e), \quad T_p(u)\le \sum_e  T_p(u,e) \le d\max_e T_p(u,e).
\ee
Let $e=(u,v)$ be the edge that achieves the maximum.
Note that any tree $T\in T_p(u,e)$ consists of the edge $(u,v)$ and two disjoint trees
$T_1\in \calT_{p_1}(u)$ and $T_2\in \calT_{p_2}(v)$,
where $p_1+p_2=p$.
Thus we have an upper bound
\[
T_p(u,e)\le \sum_{p_1+p_2=p} T_{p_1}(u) T_{p_2}(v) \le  \sum_{p_1+p_2=p} T_{p_1}T_{p_2}.
\]
Substituting it into Eq.~(\ref{Tbound2}) and taking the maximum over $u\in \calL$ we obtain Eq.~(\ref{Tbound1}).

Define a sequence $S_1,S_2,\ldots$ such that
\be
\label{Sseq}
S_1=1, \quad S_p=d\sum_{p_1+p_2=p} S_{p_1} S_{p_2} \quad \mbox{for} \quad p\ge 2.
\ee
Clearly, $T_1=S_1=1$ and $T_2\le d=S_2$.    It follows that $T_p\le S_p$ for all $p$.
In order to derive an explicit formula for $S_p$ define a generating function
$S(x)=\sum_{p=1}^\infty S_p \, x^p$. It obeys an equation
$S(x)=dS(x)^2 +x$, which implies
\[
S(x)=\frac{1-\sqrt{1-4dx}}{2d}.
\]
Taking the derivatives one gets
\[
S_p=\frac1{p!} \left. \frac{d^p S}{dx^p}\right|_{x=0} =
-\frac{(4d)^p}{2d (p!)} \prod_{a=0}^{p-1} \left( a -\frac12 \right).
\]
It follows that
\[
S_p\le \frac{(4d)^p (p-1)!}{4d (p!)}\le \frac{(4d)^{p-1}}{p}\le (4d)^{p-1}.
\]
Summarizing,  $N_p\le T_p \le S_p \le (4d)^{p-1}$.
\qqed

\subsection{Linked cluster expansion for the ground state energy}
This subsection provides the necessary tools for computing spin-spin correlators.
A reader interested only in computing the ground state energy can safely skip it.

Let us consider a more general family of Hamiltonians for which the parameter $\eps$
may be different on different edges. Let variable $\eps_{u,v}$ be assigned to an edge $(u,v)\in \calE$.
For any subset of edges $A\subseteq \calE$ denote $\eps[A]$ a collection of variables
assigned to edges of $A$. The Hamiltonian is
\be
H(\eps[\calE])=H_0 + \sum_{(u,v)\in \calE} \eps_{u,v} V_{u,v}.
\ee
Let $E(\eps[\calE])$ be the ground state energy of $H(\eps[\calE])$.
We shall consider multivariate Taylor series for the function $E(\eps[\calE])$.
\begin{lemma}
\label{lemma:manyvariables}
The multivariate Taylor series for $E(\eps[\calE])$ at the point $\eps[\calE]=0$ converges
absolutely if $|\eps_{u,v}|\le \eps_0$ for all $(u,v)\in \calE$.
\end{lemma}
{\bf Proof:}
Let $\Omega=\{ \eps[\calE]\, : \, |\eps_{u,v}|\le 2\eps_0 \; \mbox{for all} \; (u,v)\in \calE\}$.
Let us firstly show $E(\eps[\calE])$ is analytic function of each individual variable $\eps_{u,v}$
for $\eps[\calE]\in \Omega$. Indeed, let $\tilde{\calE}$ be a set of all edges except $(u,v)$.
Define an unperturbed Hamiltonian $\tilde{H}_0=H_0+\sum_{(u,v)\in \tilde{\calE}} \eps_{u,v} V_{u,v}$
and a perturbation $\eps_{u,v}\, V_{u,v}$.
It follows from Corollary~\ref{corollary:Delta/2} that $\tilde{H_0}$ has
non-degenerate ground state and the spectral gap at least $\Delta/2$. Applying the standard
perturbation theory to a perturbed Hamiltonian $\tilde{H}_0+ \eps_{u,v}\, V_{u,v}$
we conclude that $E(\eps[\calE])$ is analytic function of $\eps_{u,v}$ as long as the Weyl
condition $\| \eps_{u,v} \, V_{u,v} \|< \Delta/4$ is satisfied. Since we assumed that $|\eps_{u,v}|\le 2\eps_0$,
one has  $\| \eps_{u,v} \, V_{u,v} \|< 2\eps_0 J =2^{-17}\Delta/d < \Delta/4$.
Thus $E(\eps[\calE])$ is analytic in $\Omega$ with respect to each individual variable $\eps_{u,v}$.
Repeatedly using Cauchy's formula one gets
\be
\label{multipleCauchy}
E(\eps[\calE])= \left( \prod_{(u,v)\in \calE} \frac1{2\pi i} \oint_{|z_{u,v}|=2\eps_0}
\frac{1}{(z_{u,v}-\eps_{u,v})}\right)E(z[\calE]).
\ee
Since $H_0$ and $V$ are bounded operators, the absolute value $|E(z[\calE])|$ can be
bounded by a constant (maybe depending on $n$).
The Taylor series in $\eps_{u,v}$ at the point $\eps_{u,v}=0$
for any factor  $1/(z_{u,v}-\eps_{u,v})$ in Eq.~(\ref{multipleCauchy})
converges absolutely as long as $|\eps_{u,v}|<2\eps_0$.
Thus the Taylor series for $E(\eps[\calE])$ converges absolutely if
$|\eps_{u,v}|\le \eps_0$ for all $(u,v)\in \calE$.
\qqed

The Taylor series for $E(\eps[\calE])$ can be uniquely written in the form
\be
\label{Elce}
E(\eps[\calE])
=\sum_{A\subseteq \calE} \left(\prod_{(u,v)\in A}  \eps_{u,v} \right) p_A(\eps[A]),
\ee
where  the sum is over all subsets of edges $A$ and
 $p_A(\eps[A])$ is the series that involves only  variables $\eps_{u,v}$
pertaining to $A$. Clearly, the coefficients of $p_A(\eps[A])$ are functionals
of interactions $V_{u,v}$ with $(u,v)\in A$ only.
 The main goal of this section is to show that the
expansion Eq.~(\ref{Elce}) involves only linked clusters of edges. Let us firstly define
this notion.
 \begin{dfn}
A subset of edges $A\subseteq \calE$ is called a linked cluster iff
the subset of vertices induced by $A$ is a linked cluster.
\end{dfn}
\begin{lemma}
 \label{lemma:lce}
The series Eq.~(\ref{Elce}) involves only linked clusters of edges  $A$.
\end{lemma}
\noindent
{\bf Proof:}
Suppose  $A\subseteq \calE$ is not a linked cluster of edges.
 Let $M\subseteq \calL$ be a set of vertices induced by $A$.
Since $M$ is not a linked cluster, it can be represented as  a disjoint union
$M=M_1\cup M_2$, where $M_1,M_2\subseteq \calL$,
$M_1\cap M_2=\emptyset$, and  no edge connects $M_1$ and $M_2$.
Accordingly,  $A$ can be represented as a union $A=A_1\cup A_2$,
where $A_1$ and $A_2$ are the set of edges inducing $M_1$ and $M_2$
respectively.
Let us choose variables $\eps[\calE]$ such that
$\eps_{u,v}=0$ unless $(u,v)\in A$.
Then it is clear that the Hamiltonian $H(\eps[\calE])$ splits into a
sum of  three terms acting on non-overlapping sets of qubits:
 \[
 H(\eps[\calE])=H_1+H_2+H_{else}, \quad
 H_j=\sum_{u\in  M_j} \Delta_u \, |1\ra\la 1|_u + \sum_{(u,v)\in A_j} \eps_{u,v} V_{u,v},
 \quad
 H_{else}=\sum_{u\in \calL\backslash (M_1\cup M_2)} \Delta_u \, |1\ra\la 1|_u.
 \]
The ground state energy of $H(\eps[\calE])$ is equal to the sum of ground state energies of $H_1$, $H_2$, and $H_{else}$.
It implies that
\be
\label{additivity}
E(\eps[\calE])=E(\eps[A_1]) + E(\eps[A_2]).
\ee
If we assume that  $p_A(\eps[A])\ne 0$, when $E(\eps[\calE])$ would include at least one
monomial including variables from both sets $A_1,A_2$ which contradicts to Eq.~(\ref{additivity}).
\qqed

The following implication of Lemma~\ref{lemma:lce} will simplify computation of spin-spin correlators .
\begin{corollary}
\label{corollary:partial}
Consider a Hamiltonian $H=H_0+\eps\, V$, where $V=\sum_{(u,v)\in \calE} V_{u,v}$.
Let $E(\eps)=\sum_{p=1}^\infty E_p\, \eps^p$ be the series for the ground state energy of $H$.
Suppose the interaction $V_{s,t}$ depends on a parameter $\eta$
for some edge $(s,t)\in \calE$.
Then a derivative
\[
K_p=\left. \frac{\partial E_p}{\partial \eta}\right|_{\eta=0}
\]
can be computed by setting $V_{u,v}=0$ for all edges $(u,v)$
having distance $p+1$ or greater from the edge $(s,t)$.
\end{corollary}
\noindent
{\bf Proof:} Indeed,  $E_p$ can be obtained from Eq.~(\ref{Elce})
by setting $\eps_{u,v}=\eps$ on all edges, restricting the sum to linked clusters $A$
of size at most $p$ and collecting all monomials of total degree $p$.
Clusters $A$ that do not contain the edge $(s,t)$ will not contribute to $K_p$.
Clusters $A$ that contain the edge $(s,t)$ cannot contain any edge $(u,v)$
having distance $p+1$ or greater from the edge $(s,t)$.
\qqed

\section{Computational algorithms}
\label{sec:algorithms}
In this section we describe an algorithm that takes as input a description of the Hamiltonians
$H_0$, $V$ and an integer $p$. The algorithm returns a list of coefficients
$E_1,\ldots,E_p$ in the series for the ground state energy $E(\eps)=\sum_{p=1}^\infty E_p\, \eps^p$.
The running time of the algorithm is $n\exp{(O(p))}$.
In Section~\ref{subs:cor} we describe a generalization of the algorithm
that allows one to compute spin-spin correlation functions.

\subsection{Computing the coefficients $C_p(M)$}
\label{subs:C_p(M)}
The first  part of the algorithm is to  compute the coefficients $C_q(M)$, $\emptyset\ne M\subseteq \calL$
sequentially for $q=1,\ldots,p$ using the solution of the Kirkwood-Thomas
equations, see Eqs.~(\ref{Cformal1},\ref{Cformal2}). This gives an approximate description of the
ground state.

We shall store triples $(M,q,C_q(M))$ in $n$ bins (memory registers) $B_u$ labeled by
vertices of the graph $u\in \calL$. Once a coefficient $C_q(M)$ is computed,
the triple $(M,q,C_q(M))$ is placed into
every bin $B_u$ for which $u\in M$.   From Lemma~\ref{lemma:lcprop} we learn that
$C_q(M)=0$ unless $M$ is a subset of some linked cluster $\tilde{M}$ of size at most $q+1$.
According to Lemma~\ref{lemma:counting}, the number of linked clusters $\tilde{M}$ of size
$q+1$ containing vertex $u$ is bounded by $\exp{(O(q))}$, where the coefficient in the exponent
depends only on $d$.  Each linked cluster of size $q+1$ containing vertex $u$ has $2^q$ subsets
containing vertex $u$. Thus we can bound the number of entries in the bin $B_u$
at the moment when all coefficients $C_1(M),\ldots,C_p(M)$ have been computed
as $|B_u|\le \sum_{q=1}^p  2^q  \exp{(O(q))}=
\exp{(O(p))}$.

Suppose we have already computed all non-zero coefficients $C_1(M),\ldots,C_{q-1}(M)$, $M\subseteq \calL$.
The next step is to compute coefficients $C_q(M)$ for all sets $M\subseteq \calL$ satisfying
the condition of Lemma~\ref{lemma:lcprop}, that is $|M|_c\le q+1$.
Expanding Eq.~(\ref{Cformal2}) one gets
\be
\label{sums}
C_q(M)=E_0(M)^{-1}
\sum_{(u,v)\in \calE}
\sum_{k=1}^4
\frac1{k!}
\sum_{p_1+\ldots+p_k=q-1}
\sum_{M_1,\ldots,M_k\subseteq \calL}
C_{p_1}(M_1)\cdots C_{p_k}(M_k) \,
\la \Omega|a_M \hat{a}_{M_1} \cdots \hat{a}_{M_k}(V_{u,v})|\Omega\ra.
\ee
Note that the right hand side of this equation involves only coefficients $C_{p_j}(M_j)$ that
have been  already computed.
For simplicity let us assume that
computation of any term in Eq.~(\ref{sums}) requires one unit of time\footnote{
This assumption might seem unjustified, because the precision up to which the
coefficient $C_q(M)$ must be computed depends upon $\delta$.
However, taking into account these subtleties will lead to an additional overhead
$poly(\log{n},\log{\delta^{-1}})$ which can be neglected since the algorithm has running time
$poly(n,\delta^{-1})$.}.
Denote
\be
\label{xx}
x=\la \Omega| a_M \, \hat{a}^\dag_{M_1} \cdots \hat{a}^\dag_{M_k} (V_{u,v})|\Omega\ra.
\ee
Recall, see Claim~\ref{claim:x}, that $x=0$ unless the following conditions are met:\\
(i) Each set $M_1,\ldots, M_k$ contains at least one of the vertices $u,v$.\\
(ii) $N\backslash \{u,v\} \subseteq M \subseteq N \cup \{u,v\}$, where  $N=M_1\cup \ldots \cup M_k$.\\
The property (i) implies that for a fixed edge  $(u,v)$ we can restrict the three rightmost sums in Eq.~(\ref{sums})
by taking triples $(M_j,p_j,C_{p_j}(M_j))$ either from the bin $B_u$ or from the bin $B_v$.
Therefore for a fixed $(u,v)$, the overall number of
 non-zero terms in the three rightmost sums in Eq.~(\ref{sums}) can be bounded
by $(|B_u|+|B_v|)^k\le (|B_u|+|B_v|)^4=\exp{(O(q))}$.

We shall now prove that only a small number of edges $(u,v)$ can give a non-zero
contribution to $C_q(M)$. Indeed,  there are two cases:
(1) $M\subseteq \{u,v\}$; (2) There exists $w\in M$ such that
$w\notin \{u,v\}$.
Clearly only $O(1)$ edges $(u,v)$ can lead to the case (1), so let us focus on the
case (2).
Consider any term $x$  as in Eq.~(\ref{xx}).
Properties~(i),(ii) above imply that $x=0$ unless there exists a set $M_j$ such that
$w\in M_j$ and one of the vertices $u,v$ belongs to $M_j$.
Without loss of generality, $w,u\in M_j$.
Lemma~\ref{lemma:lcprop} implies that
$C_{p_j}(M_j)=0$ unless $|M_j|_c\le p_j+1 \le q$. Therefore
the distance between $u$ and $w$ is at most $q$.
Taking into account that $|M|\le |M|_c\le q+1$,
we can bound the number of edges $(u,v)$ that can give a non-zero
contribution to $C_q(M)$ by
$|M|d^{q+1}\le
(q+1)d^{q+1}=\exp{(O(q))}$. Summarizing, the overall number of non-zero
terms in Eq.~(\ref{sums}) is $\exp{(O(q))}$.

In order to compute all non-zero coefficients $C_q(M)$ we will have to repeat
the procedure above for each subset $M$ satisfying the condition of Lemma~\ref{lemma:lcprop},
that is $|M|_c\le q+1$. By Lemma~\ref{lemma:counting},
the number of such sets is $n\exp{(O(q))}$.
Summarizing, the overall time one needs to compute all the coefficients $C_1(M),\ldots,C_p(M)$
is $n\exp{(O(p))}$.

\subsection{Computing the ground state energy}
\label{subs:E_p}
The final step of the algorithm is to compute the coefficients $E_1,\ldots,E_p$ using
Eq.~(\ref{Eformal}). This equation can be expanded as
\be
\label{Esums}
E_p=
\sum_{(u,v)\in \calE}
\sum_{k=1}^4
\frac1{k!}
\sum_{p_1+\ldots+p_k=p-1}
\sum_{M_1,\ldots,M_k\subseteq \calL}
C_{p_1}(M_1)\cdots C_{p_k}(M_k)
\la \Omega|\hat{a}_{M_1} \cdots \hat{a}_{M_k}(V_{u,v})|\Omega\ra.
\ee
Expanding the commutators in the matrix element one gets $2^k$ terms. However, only
the term in which $V_{u,v}$ is the leftmost operator gives a non-zero contribution.
Thus
\[
\la \Omega|\hat{a}_{M_1} \cdots \hat{a}_{M_k}(V_{u,v})|\Omega\ra=(-1)^k\, \la \Omega|
V_{u,v} \,  a_{M_k}^\dag  \cdots a_{M_1}^\dag |\Omega\ra.
\]
In particular, we can restrict the summation over $M_1,\ldots,M_k$ by subsets of $\{u,v\}$ only.
There are only $3$ such subsets: $\{u\}$, $\{v\}$, and $\{u,v\}$. This observation implies that
the number of terms in the rightmost sum in Eq.~(\ref{Esums}) is $O(1)$.
Since there are $O(p^3)$ partitions $p_1+\ldots+p_k=p-1$, $k\le 4$,
the overall number of terms in Eq.~(\ref{Esums}) is $O(np^3)$.

Combining the results of Subsections~\ref{subs:C_p(M)},\ref{subs:E_p} we conclude that
the overall time needed to compute the coefficients $E_1,\ldots,E_p$ scales as $n\exp{(O(p))}$.
In the above analysis  we did not keep track of the coefficients in the exponents $O(p)$.
If one computes the exact coefficient, it yields the overall  running time
$n2^{15+6\log{(d)}}$, where $\log$ stands for the base two logarithm.
Accordingly, the running time as a function of $n$ and $\delta$ scales as
\[
T(n,\delta)\sim n(n\delta^{-1})^{15+6\log{(d)}}.
\]
For example, implementing the algorithm on a 2D square lattice ($d=4$) would require
a running time $T(n,\delta)\sim n(n\delta^{-1})^{27}$, which is certainly not practical.
Note however, that the power of $n\delta^{-1}$ depends upon the ratio $|\eps|/R$,
where $R$ is the convergence radius of the series Eq.~(\ref{Eseries}).
The power  $15+6\log{(d)}$ corresponds to the most pessimistic scenario $R=2\eps_0$
(the best lower bound on the convergence radius that we can prove) and $|\eps|=\eps_0$.

\subsection{Computing spin-spin correlation functions}
\label{subs:cor}
Let $s,t\in \calL$ be any pair of vertices.
It may or may not be an edge of the graph $\calG$.
Let us add $(s,t)$ to the set of edges $\calE$ (by creating a double
edge between $s$ and $t$ if necessary). The modified graph has
maximal degree $d^*=d+1$.
 Let $O_{s,t}$ be a Hermitian operator acting non-trivially
only on qubits $s,t$.
We shall assume that $\|O_{s,t}\|\le J$.
The quantity we are interested in is the expectation value
\[
K=\frac{\la \psi|O_{s,t}|\psi\ra}{\la \psi|\psi\ra},
\]
where $|\psi\ra$ is the ground state of  $H(\eps)=H_0+\eps\, V$.
Our goal is to compute $K$ with a specified precision $\delta$.
To this end let us define a Hamiltonian
\be
H(\eps,\eta)=H_0+\eps\,  V + \eta \, O_{s,t}.
\ee
Let $E(\eps,\eta)$ be the smallest eigenvalue of $H(\eps,\eta)$.
As we know from Lemma~\ref{lemma:manyvariables}, the Taylor series
\be
\label{eps-eta}
E(\eps,\eta)=\sum_{p,q=0}^\infty E_{p,q}\, \eps^p \, \eta^q
\ee
converges absolutely for $|\eps|,|\eta|\le \eps_0^*$, where
\[
\eps_0^*=\frac{2^{-18}\, \Delta}{d^* J}=\frac{2^{-18}\, \Delta}{(d+1) J}.
\]
The Hellman-Feynman theorem asserts that
\be
\label{HF}
K=\left. \frac{\partial E(\eps,\eta)}{ \partial \eta}\right|_{\eta=0}=\sum_{p=0}^\infty E_{p,1}\, \eps^p.
\ee
Our algorithm will get an approximation to $K$ by computing a truncation of  series in Eq.~(\ref{HF}).
The following lemma provides a bound on the error resulting from the truncation.
\begin{lemma}
\label{lemma:Kerror}
Suppose $|\eps|\le \eps_0^*/(2d)$. Then
\be
\label{Kerror}
\left|
K-\sum_{q=0}^p E_{q,1}\, \eps^q
\right|
\le
2^{-16-p}\, Jd(d+1).
\ee
\end{lemma}
\noindent
{\bf Proof:}
Let us firstly prove that
\be
\label{E_{p,1}}
|E_{p,1}|\le \frac{2^{-16}\,  d^{p+1}\, \Delta}{(\eps_0^*)^{p+1}}.
\ee
Indeed, use Cauchy's formula:
\be
\label{ci1}
E_{p,1}=\frac1{(2\pi i)^2}
\oint_{|\eps|=\eps_0^*}
\oint_{|\eta|=\eps_0^*}
\frac{E(\eps,\eta) \, d\eps \, d\eta}{\eps^{p+1} \, \eta^2} .
\ee
From Lemma~\ref{lemma:chi1} we infer that $|E(\eps,\eta)|\le 2^{-16}\, n\Delta$.
However, we would like to have an upper bound independent of $n$.
To this end we employ Corollary~\ref{corollary:partial} according to which $E_{p,1}$
can be computed by restricting the Hamiltonian on $(d+1)$-neighborhood of the edge $(s,t)$.
The number of spins in this neighborhood is at most
$n^*=d^{p+1}$.
Therefore $|E(\eps,\eta)|\le 2^{-16}\, d^{p+1} \Delta$.
Substituting this bound into Eq.~(\ref{ci1}) one gets Eq.~(\ref{E_{p,1}}).
Finally, using the condition $|\eps|\le \eps_0^*/(2d)$ we bound the sum
$\sum_{q=p+1}^\infty |E_{q,1}|\, \eps^q$  as
in Eq.~(\ref{Kerror}).
\qqed

Lemma~\ref{lemma:Kerror} shows that in order to compute $K$ with an absolute error $\delta$
it is enough to compute the coefficients $E_{0,1},E_{1,1},\ldots,E_{p,1}$ in the series Eq.~(\ref{eps-eta})
with $p=\log{(\delta^{-1})}+O(1)$.

Computation of the coefficients $E_{p,1}$ requires only minor modifications of the algorithm
described in Sections~\ref{subs:C_p(M)},\ref{subs:E_p}. Indeed, consider a function
$\tilde{E}(\eps,\eta)=E(\eps,\eps\eta)$. Using the series Eq.~(\ref{eps-eta}) one gets
\[
\tilde{E}(\eps,\eta)=\sum_{r=1}^\infty \tilde{E}_r(\eta) \, \eps^r,
\quad
\tilde{E}_r(\eta)=\sum_{p+q=r} E_{p,q}\, \eta^q.
\]
In particular,
\be
\label{Ep1}
E_{p,1}=\left. \frac{\partial \tilde{E}_{p+1}(\eta)}{\partial \eta}\right|_{\eta=0}.
\ee
On the other hand, $\tilde{E}(\eps,\eta)$ is the ground state energy of a Hamiltonian
$H_0+ \eps\, (V+\eta\, O_{s,t})$. Thus we can compute the coefficients
$\tilde{E}_1(\eta),\ldots,\tilde{E}_{p+1}(\eta)$ using the already available
algorithm for the ground state energy.
Moreover, from Corollary~\ref{corollary:partial} we know that
the coefficients $E_{0,1},E_{1,1},\ldots,E_{p,1}$
can be computed by restricting the Hamiltonian to the $(p+1)$-neighborhood of the
edge $(s,t)$. Thus we can apply Theorem~\ref{thm:algorithm} with $n$
replaced by $n^*=d^{p+1}$, obtaining an algorithm with a running time
$\exp{(O(p))}$ for computing $\tilde{E}_1(\eta),\ldots,\tilde{E}_{p+1}(\eta)$.
In fact, at every step of this algorithm we have to retain only
the terms independent of $\eta$ and the terms linear in $\eta$,
see Eq.~(\ref{Ep1}). Since we have chosen  $p=\log{(\delta^{-1})}+O(1)$,
the running time of the algorithm is $poly(\delta^{-1})$.

\section{Discussion and open problems}
\label{sec:discussion}
We have proved that the ground state properties of a spin Hamiltonian with sufficiently weak interactions
between qubits can be computed efficiently. We hope that this result could  be generalized  in
several different directions. Firstly, one could try to consider more general class of unperturbed Hamiltonians $H_0$,
for example, classical Ising-like  Hamiltonians.
In addition, one could consider systems of fermionic modes rather than spins.
Secondly, one could investigate possible generalizations of the
Kirkwood-Thomas ansatz to the case of degenerate ground state. In this case the ansatz should be constructed for
an effective Hamiltonian acting on a low-energy subspace rather than for the ground state.
Results of this kind could provide a rigorous basis for perturbative derivations of
low-energy effective Hamiltonians, for example the mapping from the half-filled Hubbard model to
the Heisenberg model.
Thirdly, one could try to get a stronger lower bound on the convergence radius $R$ of the
series $E(\eps)=\sum_{p=1}^\infty E_p\, \eps^p$.
 We note that a stronger lower bound $R\ge \Delta/dJ$ can be  easily obtained for
classical Hamiltonians, when all interactions $V_{u,v}$ are diagonal in the $|0\ra,|1\ra$
basis.  Therefore, one could speculate  that in the quantum case $R$ should be
close to $\Delta/dJ$.

\section{Acknowledgments}
The authors gratefully acknowledge useful discussions with Panos Aliferis, Barbara Terhal, and Frank Verstraete.
S.B. and D.D. acknowledge support by the DTO through ARO contract number W911NF-04-C-0098,
and D.L. by the Swiss NF and the NCCR Nanoscience.

\section{Appendix A}
In this section we prove Lemma~\ref{lemma:norms}.
By definition of the norm, $\|C\|_1=\max_{u\in \calL} Y_u$, where
\[
Y_u=
\sum_{M\ni u} E_0(M)^{-1} \left|
\la \Omega| a_M \, \hat{C}_{1} \cdots \hat{C}_{k} (V)|\Omega\ra\right|.
\]
Applying the triangle inequality one can bound $Y_u$ as
\bea
\label{X_u}
Y_u
&\le&
X_u:=\sum_{M\ni u} E_0(M)^{-1} \sum_{(v,w)\in \calE } \sum_{M_1,\ldots,M_k}
\left|
\la \Omega| a_M \, \hat{a}^\dag_{M_1} \cdots \hat{a}^\dag_{M_k} (V_{v,w})|\Omega\ra\right|\,
|C_{1}(M_1)|\cdots |C_{k}(M_k)|
\eea
Here the last sum is over all non-empty subsets $M_1,\ldots,M_k\subseteq \calL$.
Claim~\ref{claim:x} allows one to restrict the summation in Eq.~(\ref{X_u}) only by
tuples $(M,M_1,\ldots,M_k,v,w)$ satisfying conditions (i),(ii). We shall
partition $X_u$ into $k+1$ (possibly overlapping) sums
that will be dealt with separately. We define $X_u^{(j)}$, $j=1,\ldots,k$ as a sum
of all terms in Eq.~(\ref{X_u}) for which $u\in M_j$. We define $X_u^{(0)}$ as a sum of all terms
 in Eq.~(\ref{X_u}) for which $u\in \{v,w\}$. In other words,
\[
X_u^{(j)}=\sum_{M\ni u} E_0(M)^{-1} \sum_{(v,w)\in \calE } \sum_{M_1,\ldots,M_k}\chi_{M_j}(u)
\left|
\la \Omega| a_M \, \hat{a}^\dag_{M_1} \cdots \hat{a}^\dag_{M_k} (V_{v,w})|\Omega\ra\right|\,
|C_{1}(M_1)|\cdots |C_{k}(M_k)|,
\]
where $\chi_{M_j}$ is the characteristic function\footnote{That is $\chi_{M_j}(u)=1$ if $u\in M_j$
and $\chi_{M_j}(u)=0$ otherwise.} of $M_j$
and
\[
X_u^{(0)}=
\sum_{M\ni u} E_0(M)^{-1} \sum_{v\, : \, (u,v)\in \calE } \sum_{M_1,\ldots,M_k}
\left|
\la \Omega| a_M \, \hat{a}^\dag_{M_1} \cdots \hat{a}^\dag_{M_k} (V_{u,v})|\Omega\ra\right|\,
|C_{1}(M_1)|\cdots |C_{k}(M_k)|
\]
Condition (ii) in Claim~\ref{claim:x} implies
that  $u\in M \subseteq N \cup \{v,w\}$, so that
each term in Eq.~(\ref{X_u}) appears at least one time in the sums $X_u^{(0)},\ldots,X_u^{(k)}$, hence
\be
\label{overallbound}
X_u\le \sum_{j=0}^k X_u^{(j)}.
\ee

\noindent
{\bf Upper bound on $X^{(j)}$, $1\le j\le k$:}
The property (i) in Claim~\ref{claim:x} implies that at least one end-point of the edge $(v,w)$ belongs to $M_j$.
W.l.o.g. $v\in M_j$. Then property (ii) implies $M_j\subseteq M\cup \{w\}$, so that $|M_j|\le 2|M|$ (recall that
$M$ is a non-empty set because $u\in M$). It gives us a bound $E_0(M)\ge \Delta |M| \ge (\Delta/2) |M_j|$.
Note also that for any fixed $M_1,\ldots,M_k$ and $v,w$ there exist at most four sets $M$ satisfying
condition (ii) of Claim~\ref{claim:x} (take $N$ and add/subtract vertices $v$ and $w$).
Therefore
\[
X_u^{(j)}\le  \frac{8}{\Delta} \max_{M}  \sum_{(v,w)\in \calE }
\sum_{M_1,\ldots,M_k} \chi_{M_j}(u) \chi_{M_j}(v) \frac1{|M_j|}
\left|
\la \Omega| a_M \, \hat{a}^\dag_{M_1} \cdots \hat{a}^\dag_{M_k} (V_{v,w})|\Omega\ra\right|\,
|C_{1}(M_1)|\cdots |C_{k}(M_k)|.
\]
Now we can bound the matrix element by $2^k J$ and add a restriction
$M_i \cap \{v,w\} \ne \emptyset$   to the summations over sets $M_i$, $i\ne j$,
see Claim~\ref{claim:x}, property~(i).
Taking into account that
\be
\label{sumM_i}
\sum_{M_i \, : \, M_i \cap \{v,w\}\ne \emptyset} |C_i(M_i)| \le
\sum_{M_i\ni v} |C_{i}(M_i)| +  \sum_{M_i\ni w} |C_{i}(M_i)| \le 2\| C_i\|_1
\ee
we arrive to
\[
X_u^{(j)}\le  \frac{2^{2k+2} J}{\Delta}  \prod_{i\ne j} \| C_i\|_1
\sum_{(v,w)\in \calE }\sum_{M_j}
\chi_{M_j}(u) \chi_{M_j}(v) \frac1{|M_j|}
|C_{j}(M_j)|.
\]
Changing the order of summations and bounding the sum over $(v,w)$ by $d|M_j|$ one gets
\[
X_u^{(j)}\le  \frac{2^{2k+2} d J}{\Delta}  \prod_{i\ne j} \| C_i\|_1
  \sum_{M_j}
\chi_{M_j}(u) |C_{j}(M_j)| \le
 \frac{2^{2k+2} d J}{\Delta}  \prod_{i=1}^k \| C_i\|_1.
 \]
Finally, Lemma~\ref{lemma:V} implies that it is enough to consider $k\le 4$, so that
\be
\label{j=1k}
\sum_{j=1}^k  X_u^{(j)}\le  \frac{2^{12} d J}{\Delta}  \prod_{j=1}^k \| C_j\|_1.
\ee

\noindent
{\bf Upper bound on $X^{(0)}$:}
Claim~\ref{claim:x} implies that for any fixed $(M_1,\ldots,M_k,v)$ there exist at most
four sets $M$  satisfying $(ii)$. Using a bound $E_0(M)\ge \Delta$ we arrive to
\[
X_u^{(0)}\le \frac{4}{\Delta} \max_{M}
 \sum_{v\, : \, (u,v)\in \calE } \sum_{M_1,\ldots,M_k}
\left|
\la \Omega| a_M \, \hat{a}^\dag_{M_1} \cdots \hat{a}^\dag_{M_k} (V_{u,v})|\Omega\ra\right|\,
|C_{1}(M_1)|\cdots |C_{k}(M_k)|.
\]
Claim~\ref{claim:x}, property~(i) allows  us to bound the matrix element by $2^k J$ and add a restriction
$M_i \cap \{u,v\} \ne \emptyset$   to the summations over sets $M_i$.
Using Eq.~(\ref{sumM_i}) we arrive to
\be
\label{x0}
X_u^{(0)}\le \frac{2^{2k+2}J}{\Delta} \prod_{j=1}^k \|C_j\|_1   \sum_{v\, : \, (u,v)\in \calE }   1 \le
\frac{2^{10}d J}{\Delta} \prod_{j=1}^k \|C_j\|_1.
\ee
Combining Eqs.~(\ref{overallbound},\ref{j=1k},\ref{x0}) we prove the upper bound Eq.~(\ref{main1}).

The second bound Eq.~(\ref{main2}) of Lemma~\ref{lemma:norms} is much easier to prove.
Applying the triangle inequality one gets
\[
|\la \Omega|\hat{C}_1\cdots \hat{C}_k(V_{u,v})|\Omega\ra | \le
\sum_{M_1,\ldots,M_k} |\la \Omega|\hat{a}^\dag_{M_1} \cdots \hat{a}^\dag_{M_k}(V_{u,v})|\Omega\ra|
|C_1(M_1)| \cdots |C_k(M_k)|.
\]
Clearly the matrix element is zero unless $M_j\subseteq \{u,v\}$ for all $j$.
Expanding the commutators one gets $2^k$ terms, but only the term in which
all creation operators stand on the right of $V_{u,v}$ gives a non-zero contribution.
Taking into account that
\[
\sum_{M_j\, :\, M_j\subseteq \{u,v\}} |C_{j}(M_j)| \le 2\|C_j\|_1,
\]
one arrives at
\[
|\la \Omega|\hat{C}_1\cdots \hat{C}_k(V_{u,v})|\Omega\ra | \le
2^k J \prod_{j=1}^k \|C_j\|_1 \le 2^4 J \prod_{j=1}^k \|C_j\|_1,
\]
where we have applied Lemma~\ref{lemma:V} to argue that $k\le 4$.

\end{document}